\documentclass[11pt,a4paper]{article}
\usepackage{jcappub}
\usepackage{color}
\usepackage{natbib}
\usepackage{subfigure}
\usepackage{verbatim}
\usepackage{graphicx}
\usepackage{bm}              
\usepackage[amssymb]{SIunits}
\usepackage{amssymb}
\usepackage{amsmath}
\usepackage{multirow}
\usepackage{tabularx}
\usepackage{booktabs}
\usepackage{float}
\usepackage[outercaption]{sidecap} 
\usepackage{threeparttable}
\usepackage{xspace}

\usepackage[utf8]{inputenc} 

\usepackage{orcidlink}

\usepackage{pifont}

\DeclareFixedFont{\ttb}{T1}{txtt}{bx}{n}{9} 
\DeclareFixedFont{\ttm}{T1}{txtt}{m}{n}{9}  

\newcommand{\planck}{{\it Planck}\xspace}

\newcommand{\ba}{\begin{eqnarray}}
\newcommand{\ea}{\end{eqnarray}}

\newcommand  \gtsim  {\lower.5ex\hbox{$\; \buildrel > \over \sim \;$}}
\newcommand  \ltsim  {\lower.5ex\hbox{$\; \buildrel < \over \sim \;$}}

\newcommand{\be}{\begin{equation}}
\newcommand{\ee}{\end{equation}}

\bibpunct{(}{)}{;}{a}{}{,}

\def\3he{$^3{\rm He}$}

\notoc

\title{The Simons Observatory: forecasted constraints on primordial gravitational waves with the expanded array of Small Aperture Telescopes}

\collaborationImg{\includegraphics[width=2.0cm]{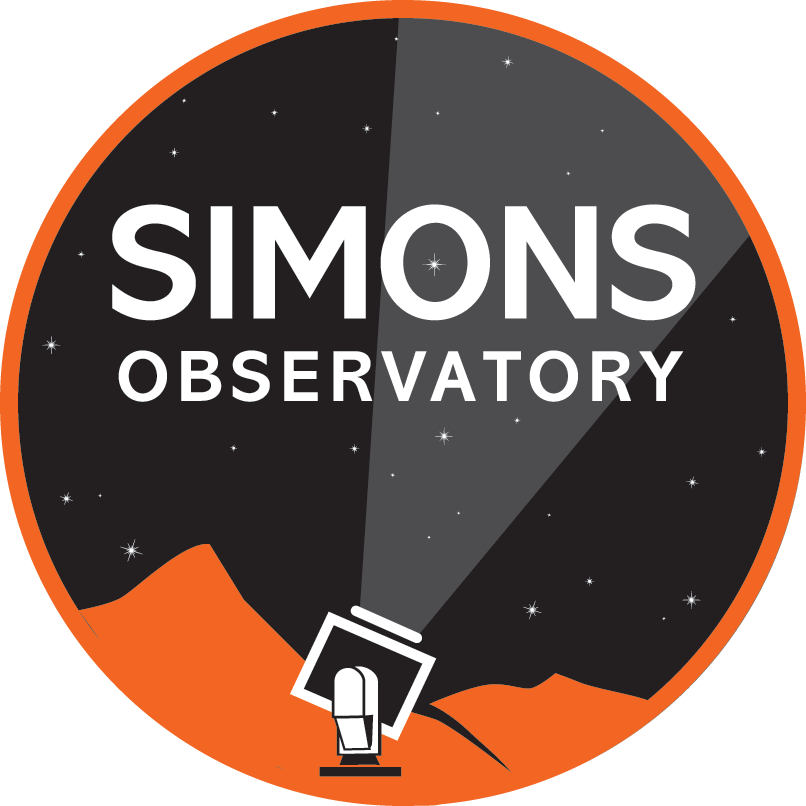}}
\vspace{10mm}
\collaboration{The Simons Observatory Collaboration}
\emailAdd{so\_publications@simonsobservatory.org}

\author[1,2]{I.~Abril-Cabezas\orcidlink{0000-0003-3230-4589},}
\author[3,4]{S.~Adachi\orcidlink{0000-0002-0400-7555},}
\author[5]{P.~Ade,}
\author[6,7,8]{A.~E.~Adler\orcidlink{0000-0002-5736-5524},}
\author[9]{P.~Agrawal\orcidlink{0009-0006-8601-6696},}
\author[10]{J.~Aguirre\orcidlink{0000-0002-4810-666X},}
\author[11]{S.~Aiola\orcidlink{0000-0002-1035-1854},}
\author[12,13]{T.~Alford\orcidlink{0000-0003-1942-1334},}
\author[6]{A.~Ali,}
\author[14]{D.~Alonso\orcidlink{0000-0002-4598-9719},}
% ...

\affiliation[1]{DAMTP, Centre for Mathematical Sciences, University of Cambridge, Wilberforce Road, Cambridge CB3 OWA, UK} 
\affiliation[2]{Kavli Institute for Cosmology Cambridge, Madingley Road, Cambridge CB3 0HA, UK} 
\affiliation[3]{Okayama University, Department of Physics, Okayama 700-8530, Japan} 
\affiliation[4]{Kavli IPMU (WPI), UTIAS, The University of Tokyo, Kashiwa, Chiba 277-8583, Japan} 
\affiliation[5]{School of Physics and Astronomy, Cardiff University, UK} 
\affiliation[6]{Department of Physics, University of California Berkeley, Berkeley, CA, USA} 
\affiliation[7]{Computational Cosmology Center, Lawrence Berkeley National Laboratory, Berkeley, CA, USA} 
\affiliation[8]{CNRS-UCB International Research Laboratory, Centre Pierre Bin\'etruy, IRL2007, CPB-IN2P3, Berkeley, US} 
\affiliation[9]{Department of Physics, University of California San Diego, San Diego, CA, USA} 
\affiliation[10]{Department of Physics and Astronomy, University of Pennsylvania, Philadelphia, PA 19104 USA} 
\affiliation[11]{Center for Computational Astrophysics, Flatiron Institute, USA} 
\affiliation[12]{Department of Astronomy and Astrophysics, University of Chicago, 5640 South Ellis Avenue, Chicago, IL, 60637, USA} 
\affiliation[13]{University of Chicago, Department of Physics, 5720 S Ellis Ave, Chicago, IL, 60637, USA} 
\affiliation[14]{Department of Physics, University of Oxford, Denys Wilkinson Building, Keble Road, Oxford OX1 3RH, United Kingdom} 
% ...

\abstract{
We present updated forecasts for the scientific performance of the degree-scale ($0.5$ deg FWHM at 93 GHz), deep-field survey to be conducted by the Simons Observatory (SO). By 2027, the SO Small Aperture Telescope (SAT) complement will be doubled from three to six telescopes, including a doubling of the detector count in the 93 GHz and 145 GHz channels to  48,160 detectors. Combined with a planned extension of the survey duration to 2035, this expansion will significantly enhance SO's search for a $B$-mode signal in the polarisation of the cosmic microwave background, a potential signature of gravitational waves produced in the very early Universe. 
Assuming a $1/f$ noise model with knee multipole $\ell_{\rm knee} = 50$ and a moderately complex model for Galactic foregrounds, we forecast a $1\sigma$ (or 68\% confidence level) constraint on the tensor-to-scalar ratio $r$ of $\sigma_r = 1.2\times10^{-3}$, assuming no primordial $B$-modes are present. This forecast assumes that $70\%$ of the $B$-mode lensing signal can ultimately be removed using high resolution observations from the SO Large Aperture Telescope (LAT) and overlapping large-scale structure surveys. For more optimistic assumptions regarding foregrounds and noise, and assuming the same level of delensing, this forecast constraint improves to  $\sigma_r = 7\times10^{-4}$. These forecasts represent a major improvement in SO's constraining power, being a factor of around 2.5 times better than what could be achieved with the originally planned campaign, which assumed the existing three SATs would conduct a five-year survey.}

\toccontinuoustrue
\begin{document}
\maketitle

\section{Introduction}
\label{sec:intro}

The Simons Observatory (SO) is a new millimetre-wave (mm) observatory that has recently started scientific operations from its high-altitude site in the Atacama Desert of northern Chile. SO currently consists of one Large Aperture Telescope (LAT, a 6 m diameter reflecting telescope) and three Small Aperture Telescopes (SATs, each consisting of a 0.42 m refractive aperture). The scientific reach of SO is wide ranging, and a comprehensive overview is presented in \cite{so_forecast:2019}. Detailed descriptions of the instrumentation already deployed on the SO site are presented in \cite{xu21}, \cite{zhu21}, \cite{galitzki24}, and \cite{2025ApJS..279...34B}. 

The SO programme is being enhanced in several ways beyond the capabilities described in~\cite{so_forecast:2019}. This paper focuses on the 0.42 m SATs and the science enabled via the degree-scale, deep survey that they are undertaking. In this regard, the major improvement programme, which is underway and already well advanced, consists of doubling the number of SATs from three to six telescopes. The other SO enhancements are described in \cite{so_lat_white_paper25}. 

As described in \cite{so_forecast:2019}, the SO SATs are designed to pursue a key science goal --- the search for a $B$-mode polarisation signal on large angular scales in the cosmic microwave background (CMB) radiation. A detection of this signal would provide compelling evidence that primordial gravitational waves were generated in the very early universe \citep{kamionkowski97, seljak97}. This search for ``primordial $B$-modes" is amongst the most prominent goals of modern observational cosmology and, in addition to SO, is being pursued by multiple ground-based, balloon, and satellite experiments \citep[e.g.][]{class_pwg16, bicep_pwg18, 2020JLTP..199..482B, lspe_pwg21, spider_pwg22, alicpt_pwg22, LiteBIRD_pwg23, 2025arXiv250502827Z}.

In this paper, we present an update to the constraints that SO is forecasted to achieve on the tensor-to-scalar ratio, $r$, taking into account the expansion of the SAT complement from three to six telescopes, and an extension of the survey duration to ten years. To produce these forecasts, we have adopted the same assumptions regarding key instrument parameters as were assumed in our previous forecasting work \citep{so_forecast:2019}. In particular, we have used the same per-detector white noise levels and the same full-width at half maximum (FWHM) model instrument beam widths as used in \cite{so_forecast:2019}. These design values for the beam FWHMs and the predicted map-level noise for the expanded 10-year, six-SAT survey configuration are presented in Table~\ref{tab:sens}. The realised values for these parameters, as measured from calibration observations with the already-deployed SATs, and a comparison of the measured performance with the design parameters used here, will be presented in forthcoming publications.

The remainder of this paper is organised as follows.  In Section~\ref{sec:infrastructure}, we briefly describe the expanded SAT infrastructure and the extended deep-field survey. In Section~\ref{sec:science}, we forecast the primary scientific analyses that will be enabled by this programme.  Much of the forecasting methodology used here is similar or identical to that employed in the SO science goals and forecasts paper~\citep{so_forecast:2019} or the complementary SO Galactic science goals and forecasts paper~\citep{SO_2022_Galactic_Science}. We direct the interested reader to those papers for further details. 
In Section~\ref{sec:summary}, we conclude and discuss the outlook for SO operations and science.

\section{Expanded capabilities}
\label{sec:infrastructure}
The three SO SATs already deployed on the site include two ``mid-frequency" (MF) SATs, operating at 93 and 145 GHz, and one ``ultra-high-frequency" (UHF) SAT, operating at 225 and 280 GHz. The ongoing programme to add to this infrastructure consists of one effort to build and deploy two further MF SATs (part of the wider SO:UK project) and a second programme to build and deploy a single ``low-frequency" (LF) SAT operating at 27 and 39 GHz.\footnote{The originally envisaged SO SAT configuration included three telescope platforms (mounts) and four receivers -- two MF, one UHF and one LF. The previous plan was to exchange one of the MF receivers with the LF receiver for one year of operations. Under the enhanced SAT programme the LF receiver will be deployed on its own dedicated telescope platform and no exchange of receivers will be necessary.} The LF SAT is being delivered, in part, by the Japan-SAT (JSAT) project. The two additional MF SATs will each field 12,040 detectors (the same as the deployed MF SATs) and will thus double the mapping speed of the SO SAT facility at the central CMB frequencies of 93/145 GHz. The LF SAT will add observational capability at lower frequencies, which will be helpful for characterising and removing Galactic synchrotron emission, a prevalent foreground contaminant in CMB observations~\citep{so_forecast:2019, SO_2022_Galactic_Science}. Once completed, the three additional SATs will increase the total number of deployed detectors on SO SATs to 61,852. 

The two additional MF receivers mostly follow the design of the original MF SATs, though there are some differences. The most significant difference is that the new MF SATs will use kinetic inductance detectors (KIDs) in place of the transition edge sensors (TESs) used in all of the other SO telescopes. KIDs \citep{day2003, doyle2008} are much simpler to read out than TESs and are thus, potentially, a key enabling technology for future CMB and mm-wave facilities planning to field instruments with many thousands of detectors. The two new SO MF SATs are expected to be the first demonstration of horn-coupled, ortho-mode transducer (OMT) dichroic KIDs on a deployed instrument. The KID arrays will be complemented by simple digital readout systems, based on radio frequency system-on-chip (RFSoC) technology \citep[see e.g.][]{liu2021}. These readout units will be used in place of the SMuRF systems \citep{yu2023} used by the other SO telescopes. 
The second significant difference is the implementation of a simplified two-lens optical design in the new MF SATs instead of the three-lens system used in the other SO SATs~\citep{galitzki24}. Note that the LF SAT is based on a three-lens optical design and will use TES detectors and SMuRF readout as per the first three SO SATs.

\begin{table*}[t]
    \caption{SO Small Aperture Telescopes Survey Specifications}
    \begin{tabular}{c | c | c | c | c | c | c }
    \hline
    \hline
        Frequency & FWHM & Baseline Depth & Goal Depth & Frequency & Detector & Int. time \\
         $\mathrm{[GHz]}$ & $\mathrm{[arcmin]}$ & $[\upmu\mathrm{K} \cdot \mathrm{arcmin}]$ & $[\upmu\mathrm{K} \cdot \mathrm{arcmin}]$ & Bands & Count & $\mathrm{[SAT} \cdot \mathrm{years}]$ \\
        \hline
        \begin{tabular}{@{}c@{}}27 \\ 39 \end{tabular} & \begin{tabular}{@{}c@{}} 91 \\ 63 \end{tabular} & \begin{tabular}{@{}c@{}} 16 \\ 10 \end{tabular} & \begin{tabular}{@{}c@{}} 12 \\ 7.7 \end{tabular} & LF & \begin{tabular}{@{}c@{}} 826 \\ 826 \end{tabular} & 8 \\
        \hline
        \begin{tabular}{@{}c@{}}93 \\ 145 \end{tabular} & \begin{tabular}{@{}c@{}} 30 \\ 17 \end{tabular} & \begin{tabular}{@{}c@{}} 1.7 \\ 2.1 \end{tabular} & \begin{tabular}{@{}c@{}} 1.2 \\ 1.4 \end{tabular} & MF & \begin{tabular}{@{}c@{}} 24,080 \\ 24,080 \end{tabular} & 36.5 \\
        \hline
        \begin{tabular}{@{}c@{}}225 \\ 280 \end{tabular} & \begin{tabular}{@{}c@{}} 11\\ 9 \end{tabular} & \begin{tabular}{@{}c@{}} 5.9 \\ 15 \end{tabular} & \begin{tabular}{@{}c@{}} 3.9 \\ 9.6 \end{tabular} & UHF & \begin{tabular}{@{}c@{}} 6,020 \\ 6,020 \end{tabular} & 10 \\
        \hline
    \end{tabular}
\begin{tablenotes}
\item Design instrument parameters used for the forecasts presented in this paper. The white noise levels (``depth'') are  provided for polarisation, for the fully completed, ten-year SO SAT survey. Two sensitivity targets are presented (baseline and goal), as in~\cite{so_forecast:2019}. Note that \cite{so_forecast:2019} provided depth levels for total intensity, which are a factor $1/\sqrt{2}$ smaller than the corresponding polarisation noise levels. Besides this, and the different number of telescopes and observing time, the depth levels quoted here also account for the impact of inhomogeneous noise as described in \cite{2302.04276}. The right-most column presents the total integration time as ``SAT-years" where one SAT-year corresponds to one SAT observing for one calendar year. The noise model assumes an observing efficiency of 20\%, based on past achieved performance \citep[see ][]{so_forecast:2019}.  Note also that the additional MF SATs will make use of KIDs instead of TESs (see main text for further details).
\end{tablenotes}
\label{tab:sens}
\end{table*}

Table~\ref{tab:sens} details how the $61,\!852$ SO SAT detectors are distributed across frequencies. The table also presents the forecasted map white noise levels for an extended-duration SO SAT survey, covering approximately 10\% of the sky and projected to finish in 2035. For simplicity, these forecasted noise levels assume identical performance for both KIDs and TESs.

All three additional SATs will use existing infrastructure from the Atacama Cosmology Telescope (ACT, \citealt{2011ApJS..194...41S}) and \textsc{Polarbear}/Simons Array experiments \citep{2008AIPC.1040...66L, 2016SPIE.9914E..1HS} at the SO site. The LF SAT will be deployed on the site previously occupied by ACT, and will modify and re-use the ACT ground screen. The two additional MF SATs will be deployed on the site previously occupied by the \textsc{Polarbear} and Simons Array telescopes. These two MF SATs will be surrounded by new purpose-built ground screens. At the time of writing the construction of this additional site infrastructure is nearing completion.

All of the new components are planned to be complete by 2027 (see timeline in Fig. 1). Observations with the first two MF SATs are already underway and the UHF SAT is currently in commissioning. For the forecasts presented in this paper, we assume a cumulative survey duration of 10 years. For the first two years, we assume operation with the original three SO SATs, and for the subsequent eight years we assume operation with the full complement of six SATs. The forecasts presented in Sec.~\ref{sec:science} are based on the cumulative sensitivity of the entire survey, accounting for the sensitivity improvement at the transition between three and six SATs. Our forecasting approach follows that explained in detail in \cite{so_forecast:2019}. We refer the reader to that paper for further details. 

\begin{figure*}[t]
\center \includegraphics[width=\linewidth]{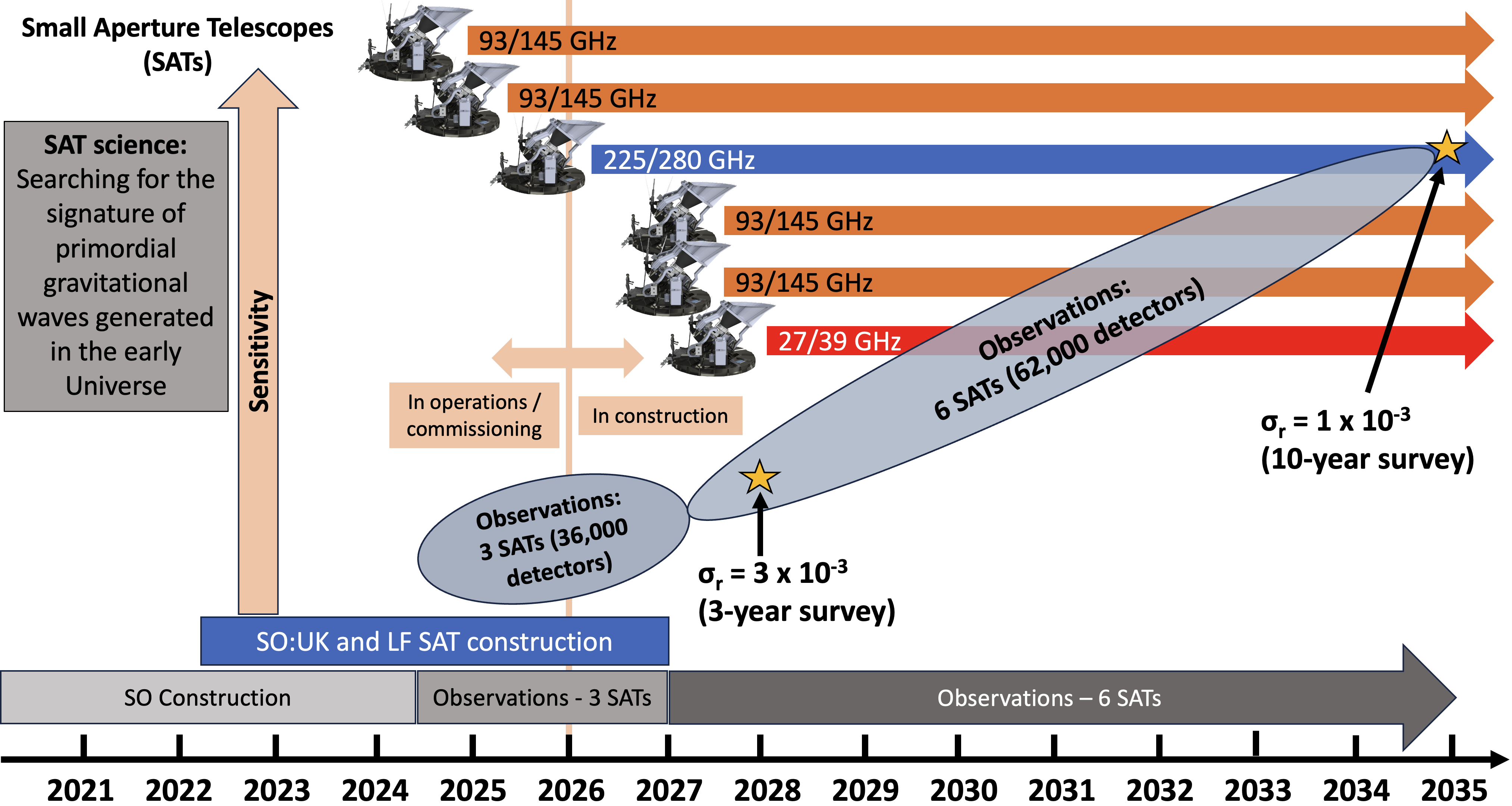}
\vspace{-5mm}
\caption{Timeline of the Simons Observatory expanded SATs programme. Two additional ``mid-frequency" SATs, operating at 93/145 GHz, are due to begin operations in 2027. A sixth ``low-frequency" SAT, operating at 27/39 GHz is due to commence operations on a similar timescale. The additional telescopes, combined with an extension of the SO SAT survey to the mid-2030s, will significantly enhance the scientific reach of the observatory. This is demonstrated here by quoting forecasts for the errors achieved on the tensor-to-scalar ratio, $\sigma_r$, by 2028 and by 2035. For clarity, forecasted errors are quoted only for the ``pessimistic" $1/f$ noise model, assuming a moderately complex model for astrophysical foregrounds (i.e. allowing for foreground frequency decorrelation -- see text in Section~\ref{sec:science} for details). See Figure~\ref{fig:sr} and Table~\ref{tab:sr} for the detailed forecasting results, covering a wider range of assumptions for the levels of noise, foreground complexity and delensing achieved.}
\label{fig:instrument} 
\end{figure*}

\section{Forecasts}
\label{sec:science}
\subsection{Forecasting framework}
\label{sec:forecast_assumptions}
To quantify the sensitivity of the upgraded SAT complement to primordial gravitational waves, and to compare it with the constraints achievable by SO before this upgrade, we carry out forecasts following the approach described in \cite{2302.04276} (W24 hereafter). More concretely, we adopt a multi-frequency power-spectrum-based component separation strategy (``Pipeline A'' in the language of W24), in which the contributions from all polarised sky components (CMB and Galactic foregrounds) are forward-modelled to the set of all auto- and cross-spectra between the 6 frequency channels observed by SO. This aligns with the main strategy used for the analysis of current SO data, as well as the approach taken by previous experiments \citep{bicep:2018}. We use only angular multipoles in the range $\ell\in[30,\,300]$, and assume a Gaussian likelihood for the measured spectra, which was found by W24 to be a good approximation in this range of scales. Further details of the pipeline implementation can be found in W24 and references therein. A validation of this approach in the presence of realistic timestream filtering and inhomogeneous noise was recently presented in \cite{2025JCAP...06..055H}. Most of the assumptions adopted for these forecasts are the same as those used in W24, in order to facilitate a comparison with previous forecasts for the nominal SO configuration.\footnote{As SO has progressed, some of these assumptions have been revised. In particular, the scanning strategy used to define the hits map used in W24 has been updated over several iterations. We have verified that, in the most stringent case (i.e. for the 10-year survey with $A_{\rm lens}=0.3$), the forecast uncertainties on $r$ reported in the next section vary by less than 10\% if analysed using the most up-to-date scanning strategy, the details of which are still under development.}

We assume the same survey strategy, resulting sky footprint and apodised mask used in W24, as well as the same noise model, characterising the contributions of white and correlated $1/f$ noise. The resulting footprint has a sky fraction of approximately $f_{\rm sky}=0.10$.\footnote{This is estimated as $f_{\rm sky}=\langle N_{\rm hits}^2\rangle^2/\langle N_{\rm hits}^4\rangle$, where $N_{\rm hits}$ is the hits count map corresponding to the selected survey strategy, and $\langle\cdot\rangle$ denotes averaging over the sky. This definition of $f_{\rm sky}$ is appropriate in the signal-dominated regime. In the noise-dominated regime, a more appropriate estimate is $f^N_{\rm sky}=\langle N_{\rm hits}\rangle^2/\langle N_{\rm hits}^2\rangle$, which is approximately $f_{\rm sky}=0.18$ for the scanning strategy assumed here.} The amplitude of the resulting noise power spectra at different frequencies are then scaled according to the sensitivity achieved by a given SAT configuration (e.g. the upgraded SAT complement after a given integration time). In particular, we assume the ``baseline'' white noise levels (see Table \ref{tab:sens}), and we produce forecasts for the two $1/f$ noise models labelled ``pessimistic" and ``optimistic" in \cite{so_forecast:2019}, and in W24. For guidance, the ``pessimistic" and ``optimistic" noise models assume knee multipole scales $\ell_{\rm knee}=50$ and $25$, respectively. We assume identical sensitivity for all four MF SATs. To produce forecasts for the error on the tensor-to-scalar ratio $r$ as a function of the total integration time, we assume the following schedule for the deployment of the different SO SATs:
\begin{itemize}
  \item At the start of the first year, the three nominal SO SATs (two of them targeting the MF bands, and one targeting the UHF bands) have been deployed.
  \item The first additional MF SAT starts observations half-way through the second year.
  \item The second additional MF SAT, and  the additional LF SAT start observations at the start of the third year.
\end{itemize}
We further assume that all telescopes will continue observations until 2035 (i.e. a total of 10 years for the nominal SATs). With this setup, at the end of the fifth year of observations, the enhanced SO SATs will have accumulated a total of 3, 16.5, and 5 telescope-years of observations in the LF, MF, and UHF channels, respectively (compared to 1, 9, and 5 for nominal SO). After the 10-year survey, the accumulated sensitivity would be 8, 36.5, and 10 telescope-calendar years, in the same channels. The noise model used assumes a 20\% observing efficiency. We emphasise that the forecasts presented in the next section depend critically on the assumed observing schedule and noise properties. They also do not account for unmodeled instrumental systematic effects. While these assumptions are based on reasonably realistic expectations, it is impossible to guarantee that they will match the final deployment schedule and instrumental performance of such a complex experiment.

As in W24, we generate a suite of simulations, each incorporating the contribution from CMB, foregrounds, and instrumental noise. The latter is both inhomogeneous and non-white (the procedure used to generate noise realisations is described in \citealt{so_forecast:2019}, and in W24). The CMB contribution to the $B$-mode signal consists solely of lensed $E$-modes, assuming no primordial $B$-modes ($r=0$). To replicate the effects of delensing, we simply scale the lensing $B$-mode power spectrum by an amplitude parameter, $A_{\rm lens}$, which quantifies the level of lensing residuals remaining in the $B$-mode power spectrum, following the application of delensing procedures. Removing approximately 50\% of the lensing $B$-mode power (i.e. $A_{\rm lens}=0.5$) should be feasible using external tracers of the large-scale structure as lensing templates \citep{1705.02332,1807.06210}. This can be further improved through ``internal'' delensing (i.e. using the lensing potential reconstructed from high-resolution CMB observations), given enough sensitivity. \cite{2110.09730} and \cite{2405.01621} showed that the nominal SO observations should be able to achieve 65\% efficiency ($A_{\rm lens}=0.35$). An extended observational campaign with the SO LAT, such as that described in \cite{so_lat_white_paper25}, would be able to improve this further. Thus, we produce forecasts for $A_{\rm lens}=0.5$ and $A_{\rm lens}=0.3$, bracketing the range of delensing efficiencies that SO could achieve, depending on the sensitivity of the LAT data. Note that we do not attempt to incorporate delensing as part of the analysis (e.g. incorporating it as part of the multi-frequency likelihood as done in \citealt{2405.01621}), but instead mimic its impact by assuming that a fraction of the lensing $B$-mode power has been subtracted from the maps. This has been shown to be a reasonable approximation for forecasting \citep{2405.01621}.

We use simple Gaussian realisations with spatially-constant spectral indices to simulate the contribution from Galactic foregrounds. These are the same Gaussian foreground simulations described in W24. This approach neglects a number of important foreground properties, including their intrinsic spatial inhomogeneity and non-Gaussianity, spatially-varying spectra, and other sources of frequency decorrelation \citep{2018MNRAS.476.1310M,2018JCAP...11..047J,2019MNRAS.487.5814V,2020ApJS..247...18W,SO_2022_Galactic_Science,2025ApJ...991...23P}. However, we incorporate the most important of these effects in the model used to fit the simulated data. This allows us to account for the impact of marginalising over the additional degrees of freedom needed to describe these effects on the final uncertainties on $r$, while ignoring the impact of varying foreground complexity on its central value, and avoiding the exploration of different foreground templates of varying complexity. As shown in W24, the impact of foregrounds on the forecast $\sigma_r$ is dominated by the level of complexity assumed in the model used for component separation, and not in the particular foreground model used to generate the simulated data (assuming unbiased constraints and adequate model complexity). A more in-depth study of foreground complexity for nominal SO is presented in W24, where the model used here was shown to be sufficiently flexible to recover unbiased constraints on $r$ in the presence of reasonably complex foreground models. See also the more recent work of \cite{2025JCAP...11..024L}. We include the contribution from Galactic synchrotron emission, assuming a power-law spectrum with spectral index $\beta_s=-3$, and Galactic polarised dust, characterised by a modified black-body spectrum with temperature $T_d=20\,{\rm K}$, and a spectral index, $\beta_d=1.54$ (\citealt{Thorne:2017}, W24). Maps for both components at their pivot frequencies ($23\,{\rm GHz}$ for synchrotron and $353\,{\rm GHz}$ for dust, respectively) were generated assuming power-law power spectra, with amplitude and spectral tilt parameters calibrated to reproduce the level of contamination present in the assumed sky footprint (see W24 for details).

We generate 500 realisations of the signal components (CMB and foregrounds), convolved at each frequency with a Gaussian instrumental beam (see Table \ref{tab:sens}). We combine them with the noise realisations described above to generate simulated observations assuming different levels of instrumental sensitivity. These noise realisations are generated directly at the map level, using the same noise power spectrum model described in W24, as well as the procedure described there to account for noise inhomogeneity. These noise curves were designed to account for the impact of time-domain filtering, and we further discard all multipoles below $\ell=30$ to avoid scales that are likely affected by atmospheric noise and aggressive filtering.

Specifically, we create simulations for the nominal SO sensitivity after 5 years of observations, as well as for the enhanced SO SAT complement every year through a 10-year observing campaign. We measure the angular power spectra of all sky realisations, and use the 500 independent simulations to construct a multi-frequency power spectrum covariance, to be used in the component separation likelihood. As in W24, we set all elements of this covariance matrix corresponding to pairs of distinct angular bandpowers to zero,\footnote{``Pairs of distinct angular bandpowers" refers to cases where the two bandpowers within a pair correspond to different multipole ranges.} in order to reduce the numerical noise in the covariance. Although this neglects the small statistical correlation within such bandpower pairs, W24 found this to be negligible. All power spectra were estimated with {\tt NaMaster} \citep{1809.09603}, making use of $B$-mode purification to eliminate the impact of $E$-to-$B$ leakage in the power spectrum uncertainties \citep{2006PhRvD..74h3002S}. We use bandpowers of width $\Delta\ell=10$, and discard all multipoles outside the range $\ell\in[30,\,300]$.

As described above, all sky components are modelled at the power spectrum level, using the model described in W24. This includes two CMB parameters, $r$ and $A_{\rm lens}$, characterising the amplitude of the tensor power spectrum, and the residual $E$-to-$B$ lensing power after delensing, respectively. In the simplest case, foreground parameters include the amplitude, spectral tilts as a function of angular scale, and spectral indices for each component, as well as a free correlation coefficient between dust and synchrotron. All these parameters, including $r$ and $A_{\rm lens}$ are free to vary in the model. To account for foreground frequency decorrelation, caused for example by spectral index variations, we use the minimal moment expansion method presented in \cite{2011.11575}, which has been shown to describe a wide range of foreground models with varying levels of complexity (\citealt{2011.11575}, W24, \citealt{2025JCAP...11..024L}). The general moment expansion approach, as proposed in the literature \citep{1701.00274,1912.09567,2411.11649} accounts for spatial variability in foreground spectra as a source of decorrelation in a perturbative manner, with the implementation used here corresponding to the minimal expansion in terms of model complexity \citep{2011.11575}. The model introduces four additional parameters, an amplitude and spectral tilt for each foreground component, characterising the spatial variability of the spectral indices for both components. When presenting our results below, the constraints obtained assuming the simpler foreground model will be labelled ``no FG decorr.'', while those using the moment expansion model to characterise frequency decorrelation will be labelled ``with FG decorr.''.

We derive constraints on all model parameters, including $A_{\rm lens}$ and $r$, by sampling the Gaussian likelihood described above using the affine-invariant Markov-chain Monte-Carlo sampler {\tt emcee} \citep{2013PASP..125..306F}. We also find the best-fit model parameters by maximising the likelihood using Powell's method \citep{powell_minim} as implemented in {\tt scipy} \citep{2020NatMe..17..261V}. We do this for a subset of 10 simulations, extracting the standard deviation on $r$ from the resulting chains, and calculating the average $\sigma_r$ across all simulations. We verify that, in all cases explored here, the best-fit value of $r$ recovered is compatible with the input value $r=0$.

\subsection{Results}
  \begin{figure}
    \centering
    \includegraphics[width=0.8\textwidth]{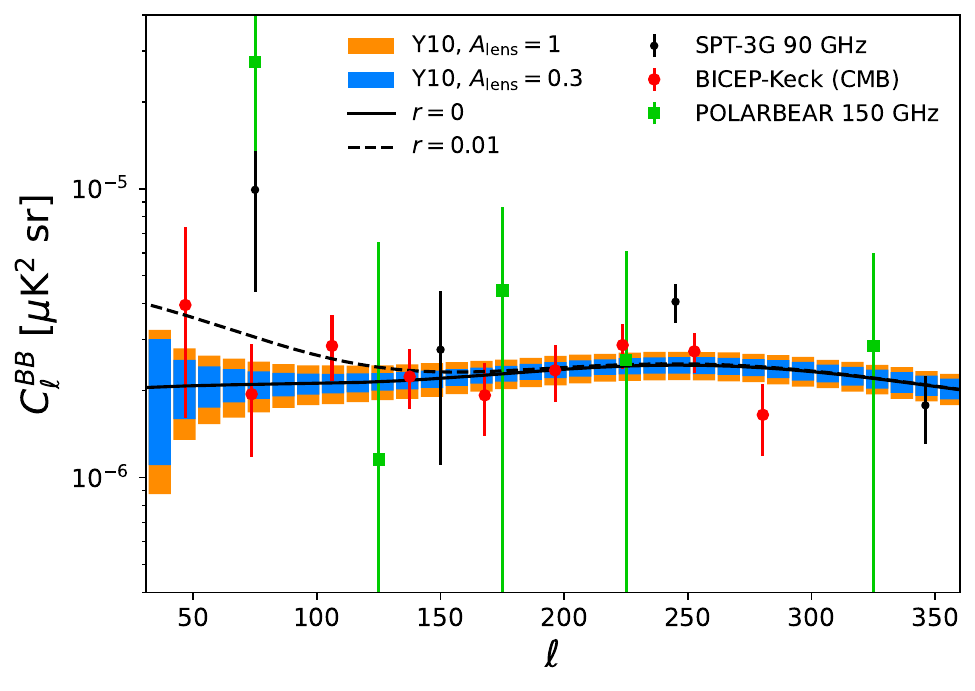}
    \caption{The $B$-mode power spectrum and uncertainties achievable by SO, accounting for the expansions described in Section~\ref{sec:infrastructure}. Bandpowers of width $\Delta\ell=10$ are shown after 10 years of observations assuming no delensing (orange bands) and 70\% delensing (blue bands), and assuming the ``pessimistic" $1/f$ noise model (see Section~\ref{sec:forecast_assumptions} for details). The solid line shows the lensing $B$-mode power spectrum in the absence of tensor modes ($r=0$). The signal for $r=0.01$ is shown in dashed black. Current measurements from BICEP/{\sl Keck} \citep[red, showing CMB bandpowers,][]{2021PhRvL.127o1301A}, SPT-3G \citep[black, showing bandpowers at 90 GHz,][]{2025arXiv250502827Z}, and \textsc{Polarbear} \citep[green, showing bandpowers at 150 GHz,][]{2022ApJ...931..101A} are also shown for reference.}
    \label{fig:dls}
  \end{figure}
  Fig. 2 shows the forecasted sensitivity of SO to the CMB $B$-mode power spectrum. The figure shows the $B$-mode bandpowers averaged over all 500 realisations used for covariance matrix estimation, together with its statistical uncertainties. Note that we do not show the measurements at a particular frequency, but rather the CMB-only bandpowers, reconstructed from the multi-frequency spectra assuming knowledge of the foreground spectral indices. The bandpower measurements are shown for the ``pessimistic'' $1/f$ noise model after 10 years of observations, assuming no delensing ($A_{\rm lens}=1$, orange bands) and 70\% delensing ($A_{\rm lens}=0.3$, blue bands). We also show the theoretical prediction for the lensing $B$-mode spectrum in the absence of tensor modes ($r=0$), as well as the signal for $r=0.01$. In this case, the primordial $B$-mode signal would be detected at high significance. For reference, the figure also shows current measurements of the $B$-mode power spectrum at ${90}\,{\rm GHz}$ from the South Pole Telescope \citep[SPT-3G,][]{2025arXiv250502827Z}, at ${150}\,{\rm GHz}$ from \textsc{Polarbear} \citep{2022ApJ...931..101A}, and the component-separated CMB bandpowers from the BICEP/{\sl Keck} collaboration \citep{2021PhRvL.127o1301A}. 
  
  \begin{figure}
    \centering
    \includegraphics[width=0.9\textwidth]{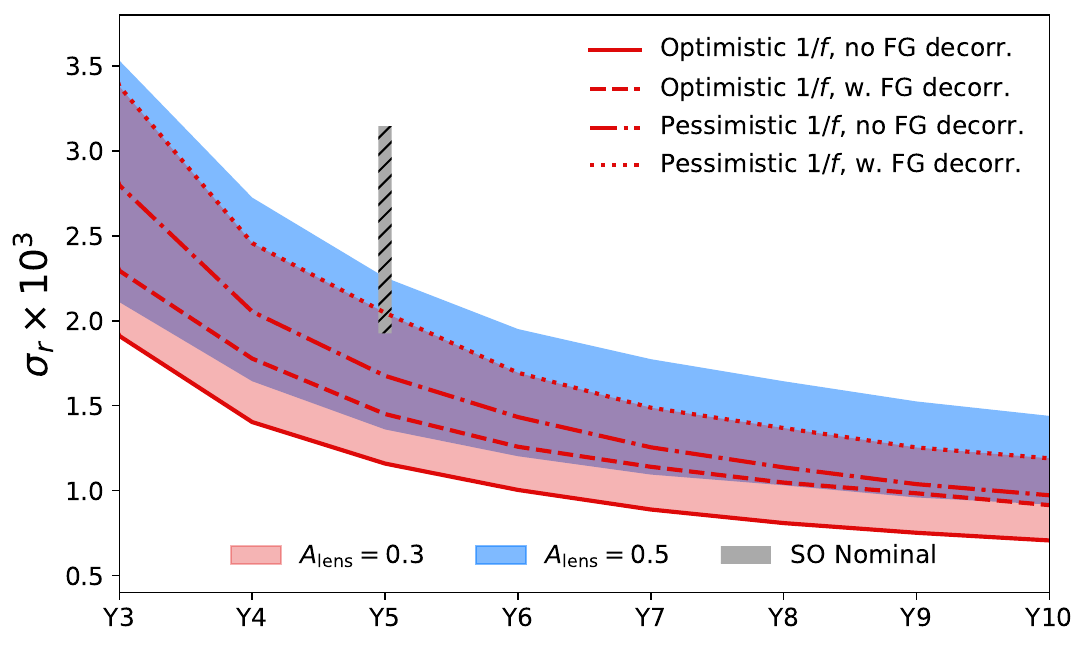}
    \caption{The forecasted 68\% constraints on the tensor-to-scalar ratio $r$, achievable by SO, as a function of time, accounting for the expansions described in Section~\ref{sec:infrastructure}. ``Y$N$'' denotes the $N$-th year after the start of observations. Results are shown assuming 50\% and 70\% delensing (blue and red, respectively), for different assumptions regarding $1/f$ noise and foreground decorrelation. Each band covers the range of $\sigma_r$ between the worst- and best-case scenarios (the ``pessimistic" $1/f$ noise model with foreground frequency decorrelation as described by the moment expansions vs. the ``optimistic" $1/f$ noise model and no decorrelation). The constraints that the nominal SO configuration would have been able to attain, with the same foreground and noise assumptions, and $A_{\rm lens}=0.5$, are shown as the gray hashed band.}
    \label{fig:sr}
  \end{figure}
  \begin{table}[]
    \centering
    \begin{tabular}{c|c|c|c|c}
    \hline
    \hline
      \multirow{2}{*}{$1/f$ noise} & \multirow{2}{*}{$A_{\rm lens}$} & \multirow{2}{*}{Survey configuration} & $10^3\times\sigma_r$ &  $10^3\times\sigma_r$ \\
       & & & no FG decorr. & with FG decorr. \\
      \hline
      \multirow{5}{*}{Pessimistic} & \multirow{3}{*}{0.5} & Nominal SO & 2.7 & 3.1 \\
                                   &                      & 5 years & 1.9 & 2.3 \\
                                   &                      & 10 years & 1.2 & 1.4 \\ \cline{2-5}
                                   & \multirow{2}{*}{0.3} & 5 years & 1.7 & 2.0 \\
                                   &                      & 10 years & 1.0 & 1.2 \\
      \hline
          \multirow{5}{*}{Optimistic}  & \multirow{3}{*}{0.5} & Nominal SO & 1.9 & 2.5 \\
                                   &                      & 5 years & 1.4 & 1.7 \\
                                   &                      & 10 years & 0.9 & 1.1 \\
                                   \cline{2-5}
                                   & \multirow{2}{*}{0.3} & 5 years & 1.2 & 1.5 \\
                                   &                      & 10 years & 0.7 & 0.9 \\
    \hline
    \end{tabular}
    \caption{Forecasted 68\% constraints on the tensor-to-scalar ratio $r$ achievable by SO, accounting for the expansions described in Section~\ref{sec:infrastructure}, after 5 and 10 years of observations. Results are shown for different assumptions regarding $1/f$ noise and the impact of foreground frequency decorrelation (as characterised by a moments expansion). For comparison, also presented are the constraints forecasted for the nominal SO configuration, without the new SATs studied here, after 5 years of observations, and assuming 50\% delensing.}
    \label{tab:sr}
  \end{table}

  Figure \ref{fig:sr} shows the forecast uncertainty on $r$ as a function of the total observing time, starting after the third year of observations. The numerical values of $\sigma_r$ presented in the plot after 5 and 10 years of observations are provided in Table \ref{tab:sr}. Results are shown assuming both 70\% and 50\% delensing, in each case presenting constraints for different combinations of assumptions regarding $1/f$ noise and the level of foreground frequency decorrelation. Frequency decorrelation plays an important role in the final constraints, which are degraded by approximately ${20}\%$ with respect to the case with no frequency decorrelation in the model. However, correlated noise has a stronger impact, with a ${30}\%$ increase in $\sigma_r$ when adopting the ``pessimistic" $1/f$ noise model. Although the impact of delensing is relatively small at first (of order a few \% improvement in $\sigma_r$ after one year in the most optimistic scenario), achieving the more ambitious 70\% delensing efficiency becomes more important towards the end of the survey, when the reduced noise levels make the impact of lensing on the final uncertainties more relevant. In this case, after a 10-year survey, going from $A_{\rm lens}=0.5$ to $A_{\rm lens}=0.3$ leads to a ${20}\%$ improvement on the final constraints. The results obtained by the nominal SO setup after 5 years, assuming $50\%$ delensing, are shown as the vertical gray hatched band, which spans the same range of scenarios as the coloured bands.

  As summarised in Table \ref{tab:sr} we find that, in the most conservative scenario that we have considered (the ``pessimistic" $1/f$ noise model, 50\% delensing, and accounting for foreground decorrelation), SO will be able to constrain the tensor-to-scalar ratio at the level of $\sigma_r=0.0014$ after 10 years of observations. We consider this most conservative scenario as our fiducial case. Forecasted constraints improve by $50\%$ to $\sigma_r=0.0007$ for the ``optimistic" $1/f$ noise model, with 70\% delensing, and assuming no decorrelation. Our fiducial constraints correspond to a ${40}\%$ improvement with respect to the bounds that would be achieved after 5 years of observations in the same scenario. These 5-year constraints are themselves 40\% tighter than those that would be achieved by the nominal SO configuration after the same observing time. Thus, in combination with the extended 10-year campaign, the upgraded SO SAT complement will improve the nominal SO constraints on primordial $B$-modes by a factor ${2.6}$, assuming 70\% delensing.

\begin{figure}
      \centering
      \includegraphics[width=0.7\linewidth]{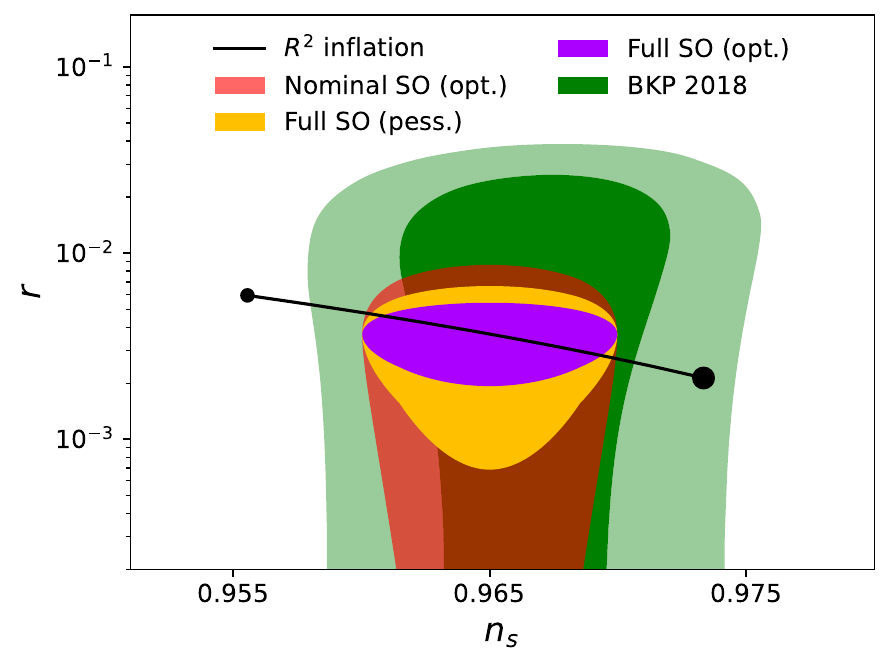}
      \caption{The forecasted 95\% constraints on the tensor-to-scalar ratio $r$ and the scalar spectral index $n_s$ that could be achieved by SO. Results are shown for the full SO SAT configuration, including the enhancements detailed in Section~\ref{sec:infrastructure}, assuming 10 years of observation and 70\% delensing. These forecasts are presented for two cases: (i) assuming the ``pessimistic" $1/f$ noise model and allowing for foreground frequency decorrelation (yellow), and (ii) assuming the ``optimistic" $1/f$ noise model and no frequency decorrelation (purple). The black line shows the prediction from $R^2$ inflation (Starobinsky model) for a total number of $e$-folds in the range $45<N_*<75$, with the range limits marked by the small and large black circles. We have assumed the true value of $r$ to correspond to the Starobinsky prediction given the best-fit value of $n_s$ preferred by \planck \citep{planck_cosmo:2018}. For comparison, the red contours show the 95\% constraints achievable by the nominal SO SAT configuration (assuming a 5-year survey and 50\% delensing, as well as optimistic $1/f$ noise and no frequency decorrelation), while the green contours show the current 68\% and 95\% constraints from BICEP/{\sl Keck}, \planck, and WMAP data \citep{2021PhRvL.127o1301A}.}
      \label{fig:inf}
  \end{figure}

  Finally, it is interesting to compare the impact these improved constraints on primordial tensor fluctuations will have on models of inflation. Fig.~\ref{fig:inf} shows the 95\% constraints achievable by SO in the $n_s$--$r$ plane, where $n_s$ is the scalar spectral index. The forecasted $n_s$ uncertainty, $\sigma(n_s)=0.002$, corresponds to the constraint achievable by the SO LAT (see \citealt{so_lat_white_paper25} for details), assuming a standard recombination history \citep{2011PhRvD..83d3513A,2011MNRAS.412..748C}. Results are shown for the full 10-year SO SAT survey (accounting for the expansions described in Section~\ref{sec:infrastructure}), for two cases: (i) assuming the ``pessimistic" $1/f$ noise model and allowing for foreground frequency decorrelation, and (ii) assuming the ``optimistic" $1/f$ noise model and no foreground frequency decorrelation. In both cases $A_{\rm lens}=0.3$ is assumed. For comparison, results are also shown for the nominal (5-year) SO SAT survey, for the ``optimistic" $1/f$ noise model, without allowing for foreground frequency decorrelation, and assuming $A_{\rm lens} = 0.5$. For illustrative purposes, we assume an inflationary signal corresponding to that predicted by Starobinsky inflation \citep{1979JETPL..30..682S} for a total of $N_*=57$ $e$-folds of inflation (which recovers the value of $n_s=0.965$ preferred by \planck (\citealt{planck_cosmo:2018}). We see that, while the nominal SO survey would not be able to detect this signal (corresponding to $r=0.0037$), the full 10-year SO SAT survey could provide evidence at the ${3}\sigma$ confidence level or higher. The figure also shows the current constraints from BICEP/{\sl Keck} \citep{2021PhRvL.127o1301A}.

\section{Summary}
\label{sec:summary}

This paper presents updated forecasts for constraints on the tensor-to-scalar ratio $r$ from the Simons Observatory. Compared to the original SO SAT configuration used for our previous forecasts \citep{so_forecast:2019}, the updated forecasts incorporate the addition of two further SATs operating in the MF bands (93/145 GHz), and the addition of one further telescope platform, which will allow continuous operation of the SO LF band (27/39 GHz) receiver until the end of the survey. In addition, an extension of the SO SAT survey duration to 2035 is considered. The two new MF SATs will exploit new detector technology, using KIDs in place of TESs, and will double the mapping speed of the SO SAT survey at 93 and 145 GHz. Continuous operation of the LF SAT will complement the frequency coverage of the experiment, providing key additional sensitivity to synchrotron contamination. These enhancements will result in a ${40}\%$ increase in the sensitivity to primordial tensor modes with respect to the nominal SO configuration after 5 years of observation, and a ${70}\%$ increase after a 10-year campaign. The final constraints on $r$ will depend critically on the level of foreground contamination, large-scale instrumental and atmospheric noise, and other instrumental systematic effects. In particular, frequency decorrelation of the Galactic foreground components can degrade final constraints by up to ${30}\%$ in the most optimistic scenario (i.e. assuming the ``optimistic" $1/f$ noise model and 70\% delensing). Given the sensitivity and sky coverage that SO will achieve, accounting for frequency decorrelation will likely be unavoidable. Likewise, large-scale $1/f$ noise has a significant impact on the final $B$-mode constraints, degrading the bounds on $r$ by $25$--$30\%$ when comparing the ``optimistic" and ``pessimistic" noise models of \cite{so_forecast:2019}. Finally, as the instrumental sensitivity grows, delensing will become increasingly necessary. After 10 years of observations, the SO constraints on $r$ may improve by around ${20}\%$ if the delensing efficiency is increased from 50\% to 70\%.

The SO SAT survey is already underway using the first two MF SATs. More recently, commissioning of the third SAT, operating at UHF (225/280 GHz), has begun. Preliminary analysis of the first data from these SATs is ongoing and initial findings on the instrument characterisation and performance will be reported in the near future. However, it is currently too early in the analysis phase to comment on the degree to which the large-scale noise properties of the observations, the complexity of Galactic foregrounds, and the levels of delensing achieved, will ultimately compare to the assumptions used for the forecasts presented here. The two additional MF SATs and the LF SAT are expected to begin operations in 2027. By this time, several SAT-years worth of survey data will have been collected with the first three SATs. In addition to delivering SO's first constraints on $r$, these early survey data will be extremely valuable for understanding the properties of the observations, and of the foregrounds in the targeted sky area. These can then be used to inform survey strategy choices for subsequent operations with the full six-telescope SO SAT configuration.

\begin{acknowledgments}
This work was supported by the United Kingdom Research and Innovation (UKRI) Infrastructure Fund and by the Science and Technology Facilities Council (STFC) (grant numbers ST/X006336/1, SMHT/X006352/1, ST/X006379/1, ST/X006344/1, ST/X006360/1, ST/X006387/1, ST/X006395/1, ST/X006328/1). This work was supported in part by a grant from the Simons Foundation (Award \#457687, B.K.). This work was supported by the U.S. National Science Foundation (Award Number: 2153201). This work was also supported by JSPS KAKENHI Grant Numbers JP22H04913, JP23H00105, JP24K23938, JP25H00403. AC acknowledges support from the STFC (grant numbers ST/W000997/1 and ST/X006387/1). GG acknowledges the support by the MUR PRIN2022 Project “BROWSEPOL: Beyond standaRd mOdel With coSmic microwavE background POLarization”-2022EJNZ53 financed by the European Union - Next Generation EU. MG, ML, PN, LP, GZ and GG acknowledge the financial support from the INFN InDark initiative and from the COSMOS network through the ASI (Italian Space Agency) Grants 2016-24-H.0 and 2016-24-H.1-2018. MG, ML, MM are funded by the European Union (ERC, RELiCS, project number 101116027). MG and MF acknowledges support from the PRIN (Progetti di ricerca di Rilevante Interesse Nazionale) number 2022WJ9J33. RB, JEG, AAK, and XR acknowledge funding from the European Union (ERC, CMBeam, 101040169) and support from the University of Iceland Research Fund. MH acknowledges support from the National Science and Technology Council and the Ministry of Education of Taiwan. The research was carried out in part at the Jet Propulsion Laboratory, California Institute of Technology, under a contract with the National Aeronautics and Space Administration (80NM0018D0004). ADH acknowledges support from the Sutton Family Chair in Science, Christianity and Cultures, from the Faculty of Arts and Science, University of Toronto, and from the Natural Sciences and Engineering Research Council of Canada (NSERC; RGPIN-2023-05014, DGECR-2023-00180). RH acknowledges support from the NSERC Canada Discovery Grant Program RGPIN-2025-06483 and the Connaught Fund. The Dunlap Institute is funded through an endowment established by the David Dunlap family and the University of Toronto. This document was prepared by Simons Observatory using the resources of the Fermi National Accelerator Laboratory (Fermilab), a U.S. Department of Energy, Office of Science, Office of High Energy Physics HEP User Facility. Fermilab is managed by Fermi Forward Discovery Group, LLC, acting under Contract No. 89243024CSC000002. CS acknowledges support from the Agencia Nacional de Investigaci\'on y Desarrollo (ANID) through Basal project FB210003. FN acknowledges funding from the European Union (ERC, POLOCALC, 101096035). CLR acknowledges support from the Australian Research Council’s Discovery Project scheme (No. DP210102386). AB, BB, SB, PC, JE, WK, WS, ETKS, and AVA were supported by the SCIPOL project,\footnote{\url{https://scipol.in2p3.fr}} funded by the European Research Council (ERC) under the European Union’s Horizon 2020 research and innovation programme (PI: J. Errard, Grant No. 101044073).
\end{acknowledgments}

\bibliography{SO-MSRI-2021,so}

\end{document}